\documentclass{article}
\usepackage[utf8]{inputenc}
\usepackage{amsmath}
\DeclareMathOperator*{\argmax}{arg\,max}

\usepackage{graphicx}
\usepackage{booktabs}
\usepackage{float}
\usepackage{amssymb}
\usepackage{csvsimple}
\usepackage{siunitx}
\usepackage{cite}
\usepackage{setspace}
\usepackage{listings}
\usepackage[a4paper, total={6in, 8in}]{geometry}
\usepackage[titletoc]{appendix}
\usepackage{xcolor}
\usepackage{soul}

\usepackage{authblk}

\DeclareSIUnit\angstrom{\text {Å}}

\title{Advanced Techniques in Automated High Resolution Scanning Transmission Electron Microscopy}
\author[1]{Alexander J. Pattison}
\author[1]{Cassio C.S. Pedroso}
\author[1,2]{Bruce E. Cohen}
\author[3,4,7]{Justin C. Ondry}
\author[3,4,5,6,7]{A. Paul Alivisatos}
\author[4]{Wolfgang Theis}
\author[8]{Peter Ercius\thanks{percius@lbl.gov}}

\affil[1]{Molecular Foundry, Lawrence Berkeley National Laboratory, 1 Cyclotron Road, Berkeley, CA, USA, 94720}
\affil[2]{Division of Molecular Biophysics \& Integrated Bioimaging, Lawrence Berkeley National Laboratory, Berkeley, CA 94720}
\affil[3]{Department of Chemistry, University of California, Berkeley, CA, USA}
\affil[4]{Kavli Energy NanoScience Institute, Berkeley, CA, USA}
\affil[5]{Material Sciences Division, Lawrence Berkeley National Laboratory, Berkeley, CA, USA}
\affil[6]{Department of Materials Science and Engineering, University of California, Berkeley, CA, USA}
\affil[7]{Department of Chemistry and Pritzker School of Molecular Engineering, University of Chicago, Chicago, Illinois 60637, United States}
\affil[8]{University of Birmingham, Edgbaston, Birmingham, B15 2TT, UK}

\date{July 2023}

\begin{document}

\maketitle
\noindent
\begin{abstract}
    Scanning transmission electron microscopy is a common tool used to study the atomic structure of materials. It is an inherently multimodal tool allowing for the simultaneous acquisition of multiple information channels. Despite its versatility, however, experimental workflows currently rely heavily on experienced human operators and can only acquire data from small regions of a sample at a time. Here, we demonstrate a flexible pipeline-based system for high-throughput acquisition of atomic-resolution structural data using an all-piezo sample stage applied to large-scale imaging of nanoparticles and multimodal data acquisition. The system is available as part of the user program of the Molecular Foundry at Lawrence Berkeley National Laboratory.
\end{abstract}

\doublespacing

\newpage

\section{Introduction}
Scanning transmission electron microscopy (STEM) is an important technique in the material sciences communities, capable of providing sub-\AA ngstrom resolution mapping of the positions of atomic columns and distributions of elements within a material \cite{Varela2005-po}. The addition of complementary modalities such as electron energy loss spectroscopy (EELS) and X-ray energy dispersive spectroscopy (EDS) has also served to increase the range and quantity of information that can be acquired through STEM. More recently, the development of high-speed pixelated direct electron detectors (DED) have made it possible to record a full convergent-beam electron diffraction pattern for every probe position in a single STEM scan, generating four-dimensional datasets (called 4D-STEM). \cite{ophus2019four} These datasets contain information about local crystallinity, determine crystal orientation \cite{Ophus2022-zp}, locate defects \cite{Mills2023-fj} and perform phase contrast imaging \cite{Chen2021-dp}.

Despite these and other developments continuing to expand the technical capabilities of STEM, the process of acquiring data has changed little since the technique's inception, at least from the operator's perspective. Although modern microscopes are (mostly) computer controlled and provide access to powerful scripting libraries, STEM experiments still generally require trained operators to be physically present to control the instrument, decide how to move the sample and choose what data to acquire. Aside from the tedium involved in such a process, this requirement a) limits access to this technique to those able to procure the services of experienced microscopist, b) limits the duration of any single microscope session to the working hours of the microscopist, c) introduces variation in the data acquisition parameters and quality and d) introduces subjective biases into resultant datasets. Additionally, the field of view of high resolution STEM is limited to tens of nanometers, making analysis of samples with micron-scale domains or features difficult. Hundreds of images with overlapping regions need to be acquired and stitched together to investigate phenomena tied to long range morphologies. Finally, human microscopists tend to expend extra electron dose when searching a sample area for features of interest by using a ``live feed" consisting of repeated consecutive images. This is of particular concern when looking at beam-sensitive samples whose structures break down under repeated imaging.

Automation is an obvious solution to this problem. Unlike humans, computers can work around the clock, perform a repetitive task consistently and operate without subjective bias (not withstanding those programmed into them). They are thus ideally suited to the acquisition of many images from a large area for statistical analysis or stitching. They can also make decisions based on single images rather than using the ``live feed" that many human operators resort to, thereby reducing the electron dose and limiting sample damage. These features are of particular interest to those working in the fields of cryo-EM and/or electron tomography, which is why many of the automation packages created for electron microscopy have been developed with one or other technique in mind. These include AutoEM and its successor AutoEMation \cite{lei2005automated}, JADAS \cite{zhang2009jadas}, JAMES \cite{marsh2007modular}, SAM \cite{shi2008script}, TOM \cite{korinek2011computer}, SerialEM \cite{mastronarde2005automated} and Leginon \cite{Suloway2005}. SerialEM and Leginon are especially popular as they are cross-compatible with Thermo Fischer (formerly FEI) and JEOL microscopes and are both open source \cite{tan2015automated}, with active developer communities dedicated to expanding their functionality in response to users \cite{Schorb2019,cheng2021leginon}. However, these programs are primarily focused on the application of conventional TEM for the biological sciences rather than STEM and the material sciences and, as such, have been optimised for a different set of experiments. While biological samples are typically far more sensitive to electron irradiation than samples studied by material scientists, the samples and set of tasks are usually also more homogeneous in nature. In cryo-EM, for example, it is quite common to combine the signal from tens of thousands of separate biomolecules to create an averaged three-dimensional model of the structure, a process known as single-particle reconstruction \cite{Nogales2016}. Here, automation is used to acquire many images of well-separated macromolecules suspended in ice with a well-established set of imaging parameters common across different macromolecules. Materials science samples are instead highly heterogeneous, where the features of interest are defects and grain boundaries whose natures are rarely reproducible or controllable. This heterogeneity has dissuaded broad adoption of automation due to the difficulty of making generalized programs. Secondly, the nature of STEM experiments (scanning a focused probe) leads to several key differences between STEM and TEM automation, especially at high resolution. These include the focusing, longer data acquisition times and the need for higher dose in general.

In this paper, we demonstrate an automated control system for STEM with a flexible control system that can be customised for specific samples and experiments. This system was designed for the TEAM 0.5 at the National Center for Electron Microscopy (NCEM) facility of the Molecular Foundry at Lawrence Berkeley National Laboratory and takes full advantage of the unique capabilities of the all piezo-electric TEAM stage. We also show integration with a fast DED showing the power of automated multimodal experimental data acquisition. All of these capabilities are offered to the international research community through the Molecular Foundry user program.

\section{Sample stage capabilities}
The TEAM Stage is an all-piezo driven stage with an ultra-low drift rate ($\sim$ 11 pm/min) and five-axis movement: $x$, $y$, $z$, $\alpha$ and $\gamma$. $\alpha$ tilt and in-plane $\gamma$ rotation provide full $\pm 180^{o}$ range of motion \cite{Ercius2012-mm}, meaning that crystalline samples can be precisely manipulated for full rotation tomography \cite{Xu2015-ef}, crystal zone-axis tilting \cite{Ben-Moshe2021-qd} and atomic electron tomography \cite{Miao2016-vx, Yang2021-vl}. The stage combines piezo motion with an accurate measurement system to implement open- and closed-loop computer control with a scripting interface. Other stages rely on sets of gears with backlash that makes accurate, repeatable motion for high resolution experiments difficult to implement. The unique properties of the TEAM Stage make it well-suited to automated and remote operation \cite{Jo2022-ru} via computer control. In this paper, we utilize the capabilities of this unique stage for automated high resolution STEM experiments.

Even so, inaccuracies in the stage motion and position measurement system limit its full potential for completing autonomous high resolution experiments. Further development in stage technology that allows for more accurate motion, repeatable motion, high stability, large range of tilts, \textit{in-situ}, and exchangeable sample magazines is required to realize the full potential of autonomous electron microscopy experiments beyond the automation demonstrated in this paper.

\section{Automation implementation}

A fully automated microscopy system needs to be able to perform, at minimum, five basic tasks to function effectively: 1) decide where to move the stage, 2) tell the stage to move, 3) check that the stage has moved to the right location (registration), 4) find the optimal focal conditions and 5) acquire data. We implemented a server/client model based on Python. A server is loaded onto the microscope PC that accepts general control commands through a networking interface, completes each assigned task (move the stage, acquire an image, change the focus, etc.) and sends results back to a local or remote client. The server is general by design and exposes all necessary microscope systems (lenses, detectors, stages, etc.) from a single program allowing for orchestration of common tasks as a queue of short sub-tasks to recover easily in case of task failure. Multiple clients can be connected from a local and/or remote PC to implement capabilities specific to a given experiment. Stage movement (task 2) and imaging (task 5) are achieved by communication with the microscope's and stage's scripting interfaces. Registration (task 3) is achieved by cross-correlation, a standard function found in most image processing libraries that can be performed quickly on modern desktop computers. More complex analysis is implemented in the client described later. The choice of the server/client model also increases the portability of the entire system. For example, a server could be written for different microscope vendors which could still communicate with a more general client.

\begin{figure} [!htb]
\centering
\includegraphics [scale=0.7] {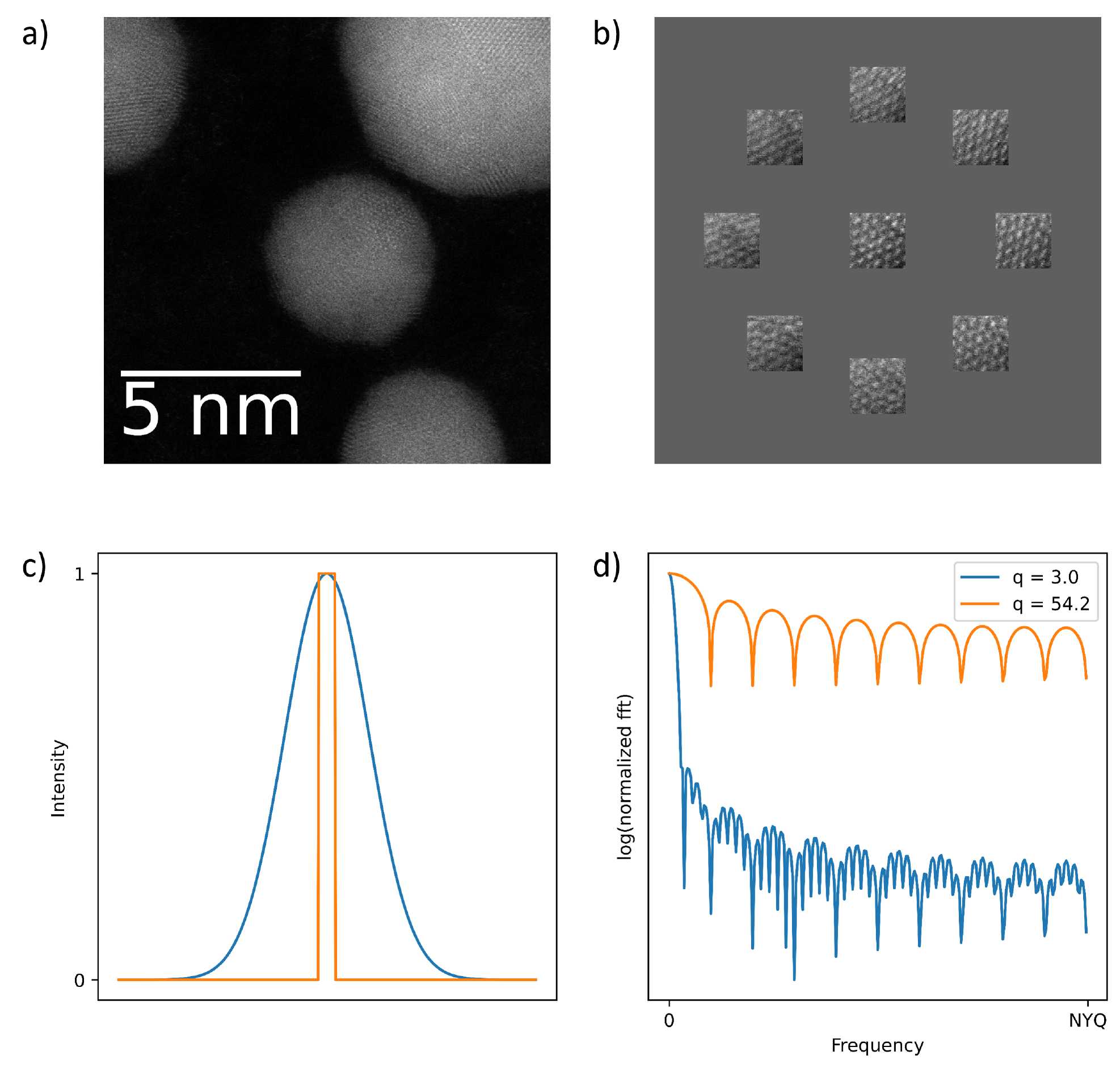}
\caption{a) HAADF-STEM image of Au NPs with b) accompanying stigmation tableau. c) Line scans and d) normalized FFTs (right) of line scans of sharp edge (orange, $q$ = 54.2) and blurred edge (blue, $q$ = 3.0). }
\label{fig:defocus slices demo}
\end{figure}

\section{Beyond Hand Panels: Assistive Functionality Towards Full Automation}

Computer control of the microscope can be used to develop new techniques that assist the user during an experiment short of full automation. A typical microscope operator station is designed around the hand panels with knobs and buttons for commonly completed tasks. Such hand panels could be considered somewhat antiquated in that they are meant for analog operation of microscope parameters by an operator with physical access to the machine. For remote operation, hand panels must be shipped to the remote site or an inefficient on-screen version is made available. Implementation of common tasks as an automated function can both improve local operation and enable remote operation. Common procedures such as object centering, focusing, and lens alignment could all be made into discrete functions for this purpose.

As an example, stigmating images is a very common practice. The user typically watches a live update of the image as the stigmator value is manually changed to achieve the best image possible. This requires significant dose, constant operator attention, and direct, analog control of the stigmator knobs. Figure \ref{fig:defocus slices demo}a-b) shows an alternative method where the computer acquires an image from a sub-region with a discrete set of stigmation values. The user (or an image quality metric) can then simply choose the best image. Such a ``stigmation tableau" provides a defined task with control over how accurate the stigmation steps should be and provides for remote/automated interaction for a very common task. This concept can be applied to almost any operational task (removal of the hand panel control) to take full advantage of computer-aided microscope control.

\section{Autofocusing}

Autofocusing (task 4) is one of the more difficult tasks in high-resolution STEM imaging due to several complications. Firstly, the complexity and heterogeneous nature of materials science samples makes development of a general algorithm difficult. Secondly, the electron beam is highly converged and very sensitive to the focus value, especially for atomic resolution imaging. Accuracy of a few nanometers is needed for the ideal imaging conditions. Optimized autofocusing routines that are fast and accurate are highly desirable for automated STEM experiments. We have developed two autofocusing routines that are applicable in a wide range of cases.

\subsection{One-dimensional intensity line cuts}

A common method of autofocusing in STEM is to use 10 to 15 scanned images of approximately 1 second each at different focii where the image quality is determined by auto-correlation. This method is fairly slow, requires high quality images to reduce noise and adds large amounts of dose to the entire sample area of interest. We instead implemented an efficient approach for STEM by analysing the sharpness of intensity data from a focal series of one-dimensional line scans taken at the same position. One-dimensional scans reduce the time needed by the square of the number of image pixels and only doses a very small region of the sample. The sharpness (quality factor) is determined for each line scan by summing the magnitudes of their fast Fourier transforms (FFT) and dividing by the magnitude of the zero frequency. The FFT of a sharp edge will have uniform magnitude at all frequencies, but a blurred edge will present suppressed higher frequencies and a lower sum, as demonstrated in Figure \ref{fig:defocus slices demo}c-d). This normalized sum of frequencies, $q$, is a useful metric of focus. In practice, the system only requires about 5 to 10 line cuts in real experiments (approximately 1 millisecond per line cut) to accurately focus, significantly reducing the time and dose required.

\subsection{Bayesian optimization}

The accuracy of line cuts is limited by the number of one-dimensional line scans according to pre-selected parameters, and it can fail if they are taken at a weakly scattering position on the sample. We have found that for atomic resolution STEM images the image quality tends to be highly peaked requiring several line-cut iterations with reduced step size at each stage to achieve the best focus. Thus, the line scan method works best on samples with very high contrast like well separated nanoparticles and less well on continuous thin films. Lastly, parallel beam TEM can only be acquired as full images, and this method can not be used.

Thus, in some cases it is better to use full images to achieve the best focus, and an algorithm that minimizes the number of full images is desirable. We have implemented a Bayesian optimization technique to auto-focusing. Bayesian optimization is an efficient method of estimating expensive-to-compute unknown functions by minimizing uncertainty \cite{shahriari2015taking} and is being used for autonomous experiments in synchrotrons \cite{noack2021gaussian}, fMRI studies \cite{lorenz2017neuroadaptive} and scanning-probe microscopy \cite{thomas2022autonomous}. It allows us to use the confidence in the peak value of a scalar image quality metric to determine when the optimal focus has been found. We can also adapt the algorithm to achieve higher accuracy (at the cost of more images/dose) on radiation-hard materials or lower accuracy (reducing dose) on dose sensitive materials.

Bayesian optimization requires a scalar metric of focus quality. Kirkland demonstrated in simulation that normalized image variance is a good, simple indicator of the focal quality of an atomic resolution high-angle annular dark field (HAADF-) STEM image. \cite{kirkland1990image} Normalized image variance is calculated as

\begin{equation}
    F(df) = \frac{\sigma^2}{\mu^2} = MAX
\end{equation}

\noindent
where $\sigma^2$ is the variance of the pixel intensities of an image and $\mu$ is the mean pixel intensity, such that

\begin{equation}
    \mu = \frac{1}{N} \sum_{ij} z_{ij}
\end{equation}

\begin{equation}
    \sigma^2 = \frac{1}{N} \sum_{ij} (z_{ij}-\mu)^2
\end{equation}

\noindent
where $z_{ij}$ is the pixel intensity at position $i, j$ and $N$ is the total number of pixels \cite{kirkland2018fine}.

In STEM, while acquiring an image and calculating the variance are simple operations, each image taken increases the applied dose and the time required ($\sim$1-3 seconds) for high-quality scanned images at atomic resolution is non-trivial. Thus, taking images from which to evaluate focal quality can be considered expensive in terms of a long-running automation experiment where auto-focusing is used in multiple steps. In Bayesian optimization, a surrogate model is generated as a proxy for the black-box function, usually using Gaussian processes (GP) \cite{shahriari2015taking}. Bayes' Theorem is then used to incorporate prior (and subsequent posterior) knowledge in order to select the most beneficial measurement value (i.e. focus) to acquire next. This selection is determined by an acquisition function, the choice of which determines the behaviour of the optimization routine \cite{shahriari2015taking}. A common choice for determining the maximum of a function is the upper confidence bound (UCB) method, where the choice of the next measurement point is determined by the maximum of the UCB. This process can be mathematically represented by:

\begin{equation}
    x^* = \argmax{[f(x)+A\times\sigma(x)]}
\end{equation}

where $x^*$ is the next sample point to be chosen, $f(x)$ is the posterior focus quality at defocus setting $x$, $\sigma(x)$ is the posterior uncertainty of the posterior focus quality at defocus setting $x$ and $A$ is a hyperparameter that determines the size of the confidence bound \cite{shahriari2015taking}. Increasing $A$ favors exploration (searching new regions to try to find the maximum) while decreasing $A$ favors exploitation (searching close to the previously discovered maximum).

In this work, a Bayesian optimization routine using Gaussian processes and the UCB acquisition function was implemented using the Gaussian process module in \textit{scikit-learn} version 1.0.2 \cite{scikit-learn}. Since the relationship between defocus and image variance of a high resolution image approximates a smooth Gaussian with a single peak, $A$ was set to 1.5 to more strongly favor exploitation rather than exploration, thereby decreasing the number of acquisitions required to find the maximum.

Gaussian processes are designed to explore a function space and to accurately predict the function everywhere, which is incompatible with the goal of only finding the maximum image variance (i.e. best focus). Still, the Bayesian optimization method allows us to use the confidence in the peak to define our desired accuracy in the image focus, which might require fewer or more images than what was pre-determined for i.e. a focal series. Thus, we developed an early stopping algorithm to further reduce the number of scans according to these conditions:

\begin{enumerate}
    \item Check that the maximum of the lower confidence bound, which should always be close to the global maximum of the upper confidence bound, is greater than all the local maxima of the upper confidence bound except for the global maximum. This indicates that the current peak of the surrogate function is the main peak of the surrogate function to a high degree of confidence.
    \item Check that the average distance between the last $n$ sampled defocus values is smaller than a threshold $t$. In this work, $n$ was set to 3 and $t$ to 1 nm. 3 is a good minimum for $n$, since this, in principle, corresponds to testing one value in the centre of the peak and one value either side of it. The choice of $t$ is sample-dependent, with thinner samples having narrower defocus tolerances than thicker samples.
\end{enumerate}

\begin{figure} [!htb]
\centering
\includegraphics [scale=0.5] {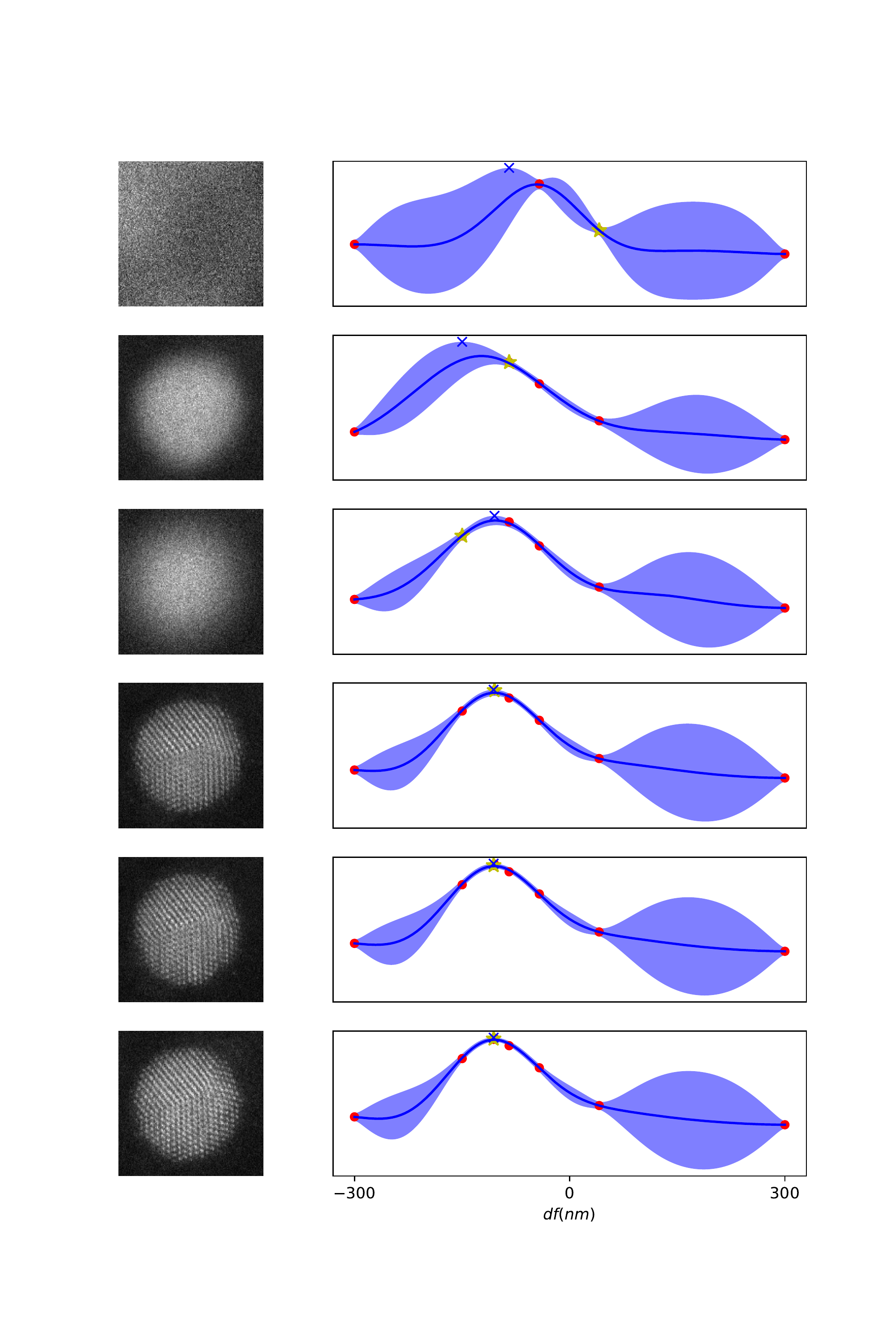}
\caption{Auto-focusing on a Au nanoparticle. The HAADF-STEM images on the left correspond to the defocus values denoted by the yellow stars on the right. The graphs on the right demonstrate the Bayesian optimization process, with the blue lines representing the surrogate model, the shaded bars representing the confidence bounds (1.5$\sigma$), the red dots denoting previously sampled defocus values, the yellow stars denoting the most recently sampled defocus value, and blue cross denoting next defocus value to be sampled. The red sample points in the first row were chosen as part of the initial data set, while all subsequent sample points were determined by the program.}
\label{fig:Exp_BO}
\end{figure}

This optimization algorithm was tested on a Au nanoparticle sample. For early stopping condition 2, $n$ was set to 3 and $t$ was set to 1 nm. The defocus was changed by 100 nm from the visually determined optimum and the optimizer was instructed to search with a boundary of $\pm$300 nm defocus. The optimizer was initialised with 3 starting points: 2 placed at the bounds and 1 chosen randomly. An example of this optimization routine is shown in Figure \ref{fig:Exp_BO}, which shows how the routine quickly and accurately determines the optimal defocus value in just six more scans. The exact number of scans needed fluctuates between runs due to the randomness of the initialization and noise. This will also vary depending on the sample and the values of the hyperparameters, such as number of starting points and the choice of $A$ in the UCB function. We also found that this technique works at high resolution on a polycrystalline thin film (not shown), making this a generalizable autofocus routine.

The ability to efficiently and accurately auto-focus on any arbitrary sample is vital to obtaining high-resolution data from automated STEM. Additionally, normalized image variance can also be used to correct non-round aberrations (astigmatism, coma, etc.) making Bayesian optimization a general tool for electron microscope alignment \cite{kirkland2018fine}. Future work will explore the use of Bayesian optimization to compensate for higher-order lens aberrations (such as astigmatism) in multiple dimensions simultaneously and line cuts to further reduce applied dose and time.

\section{Automated Workflows}

Once the common steps of STEM operation become discrete functions, an experiment can be expressed as a set of repeatable steps integrated into a workflow. Many processes (like defocus and stigmation) can be written as self-contained elements that can be reused in different parts of the workflow or in different workflows. Thus, in order to be as broadly usable as possible, our automation software program is highly modular. The workflow for each experiment consists of a pipeline of elements that can be customised and chained together to accomplish almost any task. This means that any new function developed for specific workflows can be easily reused in future experiments.

Each element consists of one of three main methods:

\begin{itemize}
  \item Tiling: This creates a grid of target sample positions at which to acquire data. The shape of the grid, spacing or overlap between adjacent tiles, and field-of-view of these tiles can all be specified by the user. This method is typically used to study continuous thin films or samples with a dense packing of nanoparticles.
  \item Point of Interest: This identifies points of interest in an overview image for the microscope to investigate. Points of interest are typically identified automatically in an initial low resolution overview image. This method is generally used for samples with a sparse dispersion of nanoparticles.
  \item Record: This contains all the parameters needed for recording data and is typically the final element in any pipeline. Available data streams are the HAADF detector and a direct electron detector.
\end{itemize}

Any number of elements can be chained together into a pipeline to perform complex tasks and then saved as a text file for reuse later. A pipeline is implemented in a workflow client that steps through each stage of the pipeline adapting to the current microscope and sample state at each step. The automation program with an example of a pipeline (left panel) that acquires a 3 by 3 grid is shown in Fig \ref{fig:HTP Client} (see Appendix \ref{sec:pipeline} for full 4D-STEM pipeline text). The pipeline is organized as a hierarchical set of elements with the associated parameters and are described below.

\begin{figure} [!htb]
\centering
\includegraphics [scale=0.5] {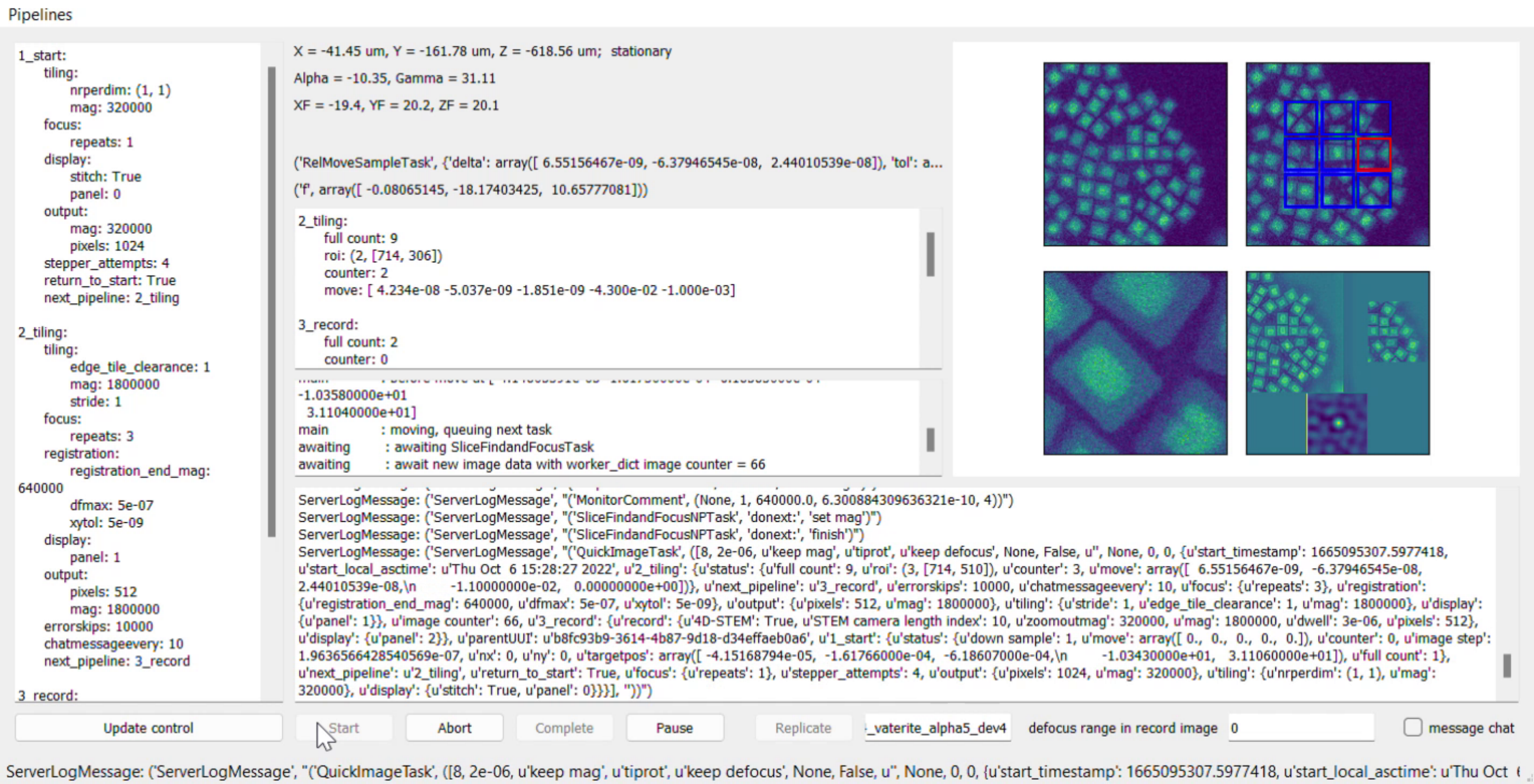}
\caption{A screenshot of the automation client shows the pipeline (left panel) of steps to be completed and the associated parameters. The current status of the machine, its location, and the relevant images/data are also displayed to the operator.}
\label{fig:HTP Client}
\end{figure}

The first element \textit{1\_start} generates a large-field-of-view reference image to be used for subsequent registration. This is done by using the tiling method and setting 'nrperdim' (number of tiles per dimensions) to 1 in each dimension. The field of view of these tiles is determined by the magnification setting (320k$\times$ in this case). 

The second element \textit{2\_tiling} generates the grid of positions for the microscope to step between and takes images. In this case, the size of the grid is calculated from the specified magnification, stride and edge tile clearance rather than being explicitly stated by the user. Stride determines the spacing between adjacent tiles, with a stride of 1 meaning that the spacing between the tile centres equals the size of the tiles (i.e. the edges of each tile touch perfectly). Edge tile clearance determines how many tiles to remove around the edge of the grid. This is done because the registration process for tiles around the edge is typically more prone to errors due to the lesser overlap between the registration image and the reference image. With a stride of 1, a 5 by 5 grid of 1.8Mx tiles would fit inside a single 320kx reference image, but the edge tile clearance of 1 reduces this to 3 by 3. Overlapping of images is desirable to create a montage with no breaks between regions but sacrifices efficiency by imaging the same region many times. Further, stitching high-resolution images becomes problematic due to drift at the atomic scale. Alternatively, a sufficient gap can be used to ensure that every image only contains new sample regions which is desirable when trying to determine morphological statistics from many objects.

The third element \textit{3\_record} contains the necessary recording parameters: dwell time, magnification, number of scan positions and camera length. After the high-resolution data acquisition, it lowers the magnification to acquire an overview useful to more easily register this image relative to the original reference image. We also implement the ability to acquire multimodal data, which in this case is a 4D-STEM data set.

Each high-level element contains additional parameters pertaining to lower-level functions. 

\begin{itemize}
    \item Focus: contains all parameters connected to auto-focusing including repeats to check accuracy, number of defocus steps and other advanced parameters. 
    \item Registration: contains all parameters that handle registration, including the final magnification at which to perform registration for this step (registration\_end\_mag), the maximum permitted value of the microscope defocus (dfmax), and the tolerance for deviation from the target position (xytol).
    \item Display: contains all parameters that determine how information is displayed in the workflow display including data stitching and which panel the output appears.
    \item Output: contains the image parameters for the output from all elements (except for the final record operation, which uses `record').
\end{itemize}

The pipeline is a hierarchical set of tasks that takes the output from a previous pipeline element and applies the next set of tasks taking into account the current state of the microscope. For example, the tiling element takes the initial overview image and generates a sequence of outputs that are passed down to the next element, which in turn generates its own set of outputs. This makes the system more robust and allows it to restart at different steps in the full process, avoiding the need to completely restart a pipeline if one failure occurs. There are also high-level parameters, such as the number of attempts to reach a position (stepper\_attempts), whether or not the stage returns to its original position after the pipeline has been completed (return\_to\_start), how many errors to tolerate before aborting (errorskips) and the ability to send updates as chat messages during operation (chatmessageevery).

\section{High-throughput automated HAADF-STEM data acquisition}

An example of the capabilities of our pipeline-based system is in the automated acquisition of atomic resolution HAADF-STEM images of CdSe/CdS core-shell nanoparticles. These particles were arranged in a highly ordered array that can be fused together to form a superlattice. Ideally, the particles in these arrays should share the same lattice orientation prior to fusing. However, making a quantitative assessment of the orientations across large regions with reliable statistics requires atomic-resolution images of thousands of particles. An automated system such as ours is ideally suited to this.

\begin{figure} [htb!]
\centering
\includegraphics [scale=0.3] {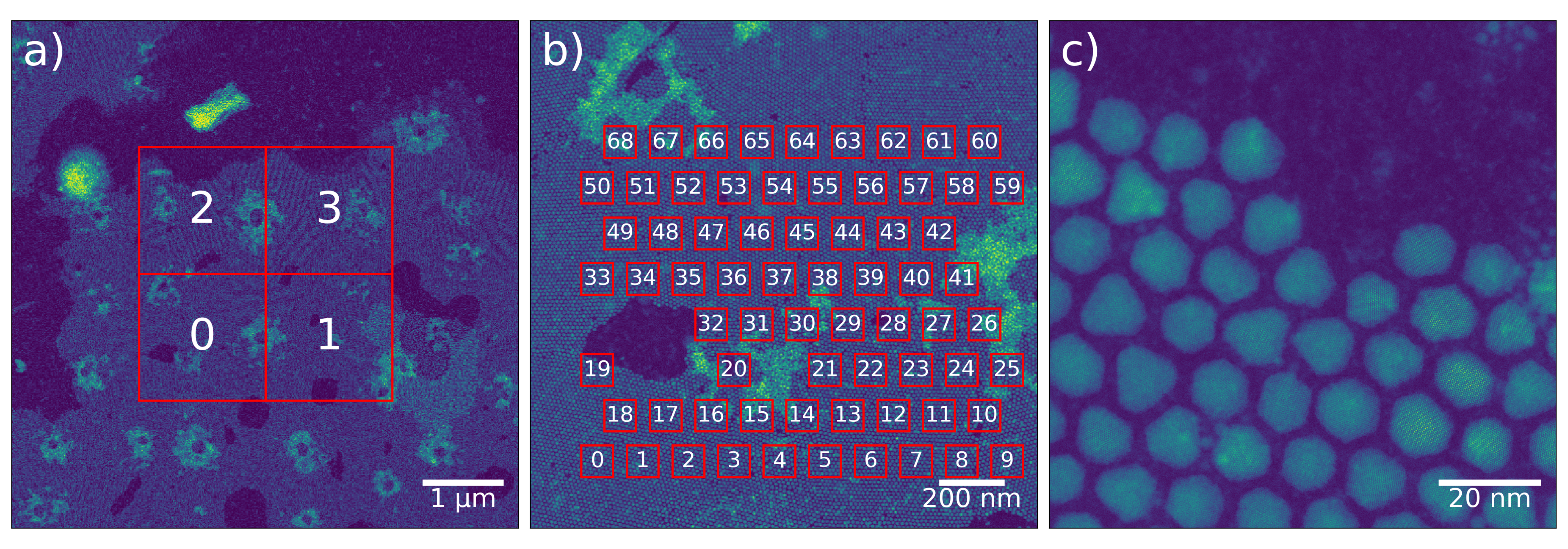}
\caption{Automated atomic resolution imaging of CdSe/CdS core/shell nanoparticles. a) Low magnification image with four regions of interests (ROIs) to be processed sequentially. b) Image after moving the stage to ROI 0, with ROIs for the atomic resolution images indicated. c) Atomic resolution image of ROI 19.}
\label{fig:CdSSe-PAE}
\end{figure}

\begin{figure} [htb!]
\centering
\includegraphics [scale=0.4] {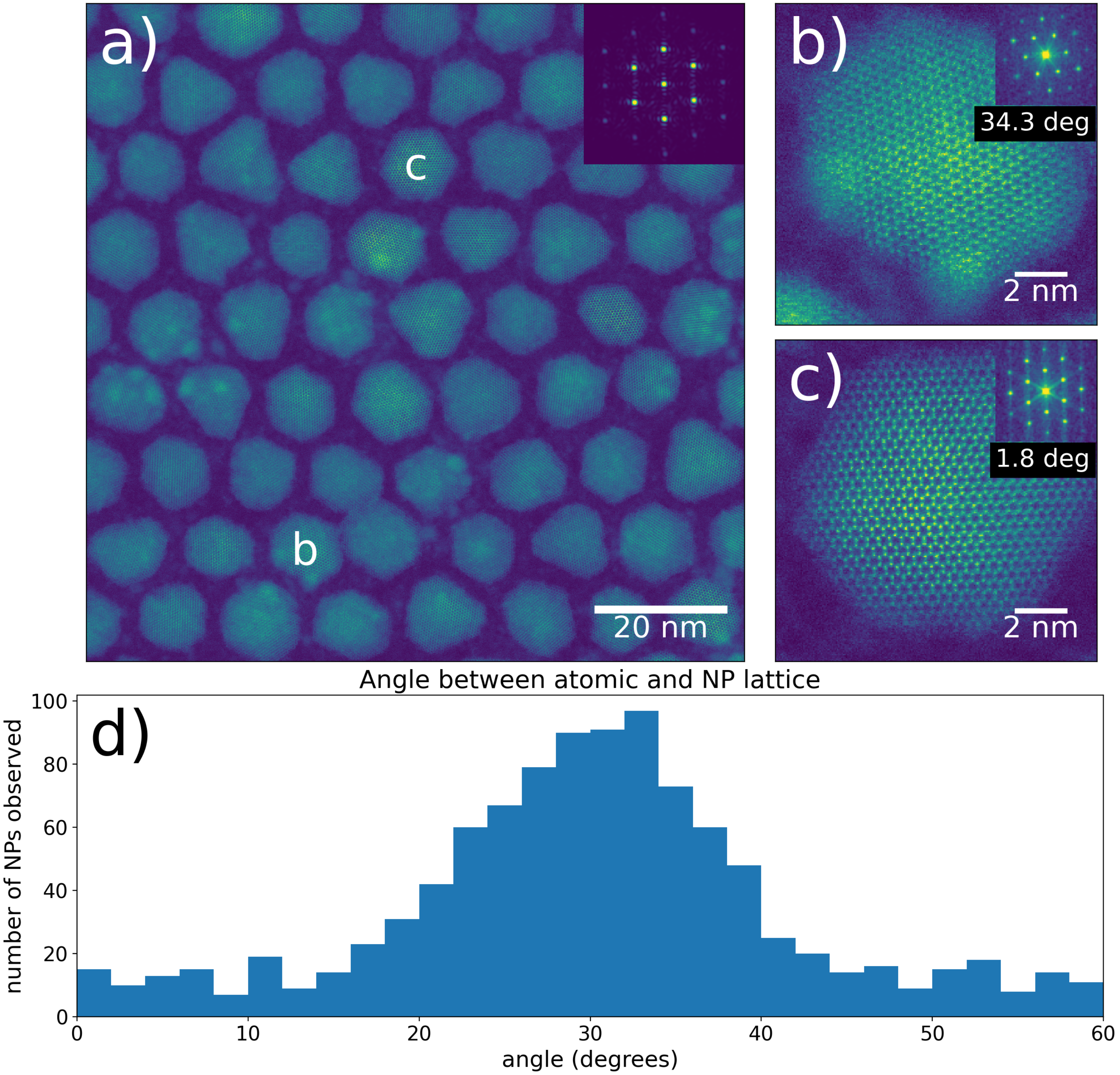}
\caption{Atomic lattice orientation of CdSe/CdS core/shell nanoparticles relative to the nanoparticle superlattice. a) Atomic-resolution image exhibiting a moderately ordered hexagonal arrangement of the nanoparticles. The inset shows the Fourier pattern of the nanoparticle positions. b-c) Cropped sections of image (a) with nanoparticles labeled as b and c, respectively. The insets are Fourier patterns of the atomic resolution images. The relative angle between atomic lattice and superlattice Fourier pattern is indicated. d) Histogram of the relative in-plane angle for the 1081 nanoparticles with the closest [100] zone-axis alignment within the set of 18,788 imaged nanoparticles.}
\label{fig:CdS-CdSe-lattice}
\end{figure}

Figure \ref{fig:CdSSe-PAE} illustrates the automated high-resolution HAADF-STEM imaging of an approximately 2.5 $\mu$m region of a CdSe/CdS core/shell superlattice.
The pipeline used (see Appendix \ref{sec:CdSSe Pipeline}) first acquired a low resolution image at 5 k$\times$ magnification to provide a starting point for the image registration of two subsequent levels of tiling followed by the capture of the respective high resolution images.
In a continuous 8.5-hour session without human intervention, 146 high resolution images were acquired, each capturing on average 40 nanoparticles at atomic resolution (see 
Figure \ref{fig:CdS-CdSe-lattice}a-c).

The superlattice was grown such that the [100] axis of the CdSe/CdS nanoparticles should be approximately aligned with the substrate normal. Combined data from several sessions comprising 18,788 nanoparticles was analyzed to elucidate the orientation of the nanoparticles' atomic lattice relative to the substrate plane and the close-packed lattice that the nanoparticles arrange in on the substrate.

Figure \ref{fig:CdS-CdSe-lattice}a) shows one of the 482 atomic-resolution images and as an inset the Fourier pattern of the nanoparticle positions. Sections of the image with nanoparticles labeled as b and c are cropped and displayed in Figure \ref{fig:CdS-CdSe-lattice}b) and c), respectively, showing a near [100] orientation. Their Fourier patterns demonstrate that they have different in-plane orientations relative to the superlattice. The large high-throughput data set allowed us to explore the relative orientation of the atomic and nanoparticle superlattice in detail. For the analysis, the nanoparticles that, based on their Fourier patterns, were highly aligned to the [100] zone axis were selected. For these 1013 nanoparticles the relative in-plane orientation of atomic lattice and superlattice were determined. As both lattices are hexagonal, the relative angle can be chosen to be modulus 60 degrees. Figure \ref{fig:CdS-CdSe-lattice}d) provides a histogram of the relative angles. It is strongly peaked at 30 degrees with a value roughly eight times as large compared to the alignment of the lattices (0 degrees). The preference for the 30-degree orientation is consistent in the context of the equilibrium facet geometry of the nanoparticles in that it offers a denser packing of the superlattice. However, many of the nanoparticles in the experiment have non-regular shapes. Thus, high-throughput data was required to elucidate the relation between the atomic lattice and superlattice orientation.

\section{Multimodal automated data acquisition}
\begin{figure} [!htb]
\centering
\includegraphics [scale=1.0] {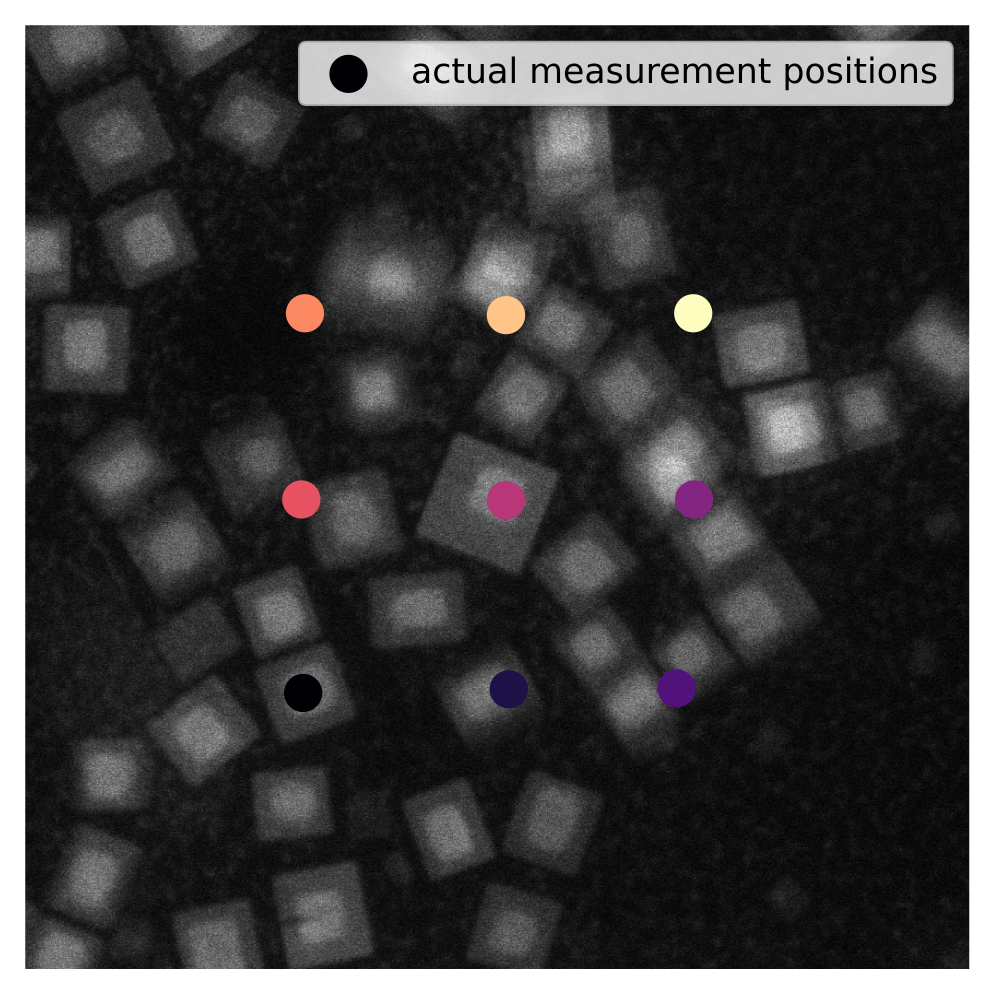}
\caption{Overview HAADF-STEM image of tiled 4D-STEM data acquisition. The colored dots indicate the positions of each data acquisition determined by cross-correlation using the post-acquisition zoom-out image.}
\label{fig:4d-pos}
\end{figure}

\begin{figure} [!htb]
\centering
\includegraphics [scale=0.45] {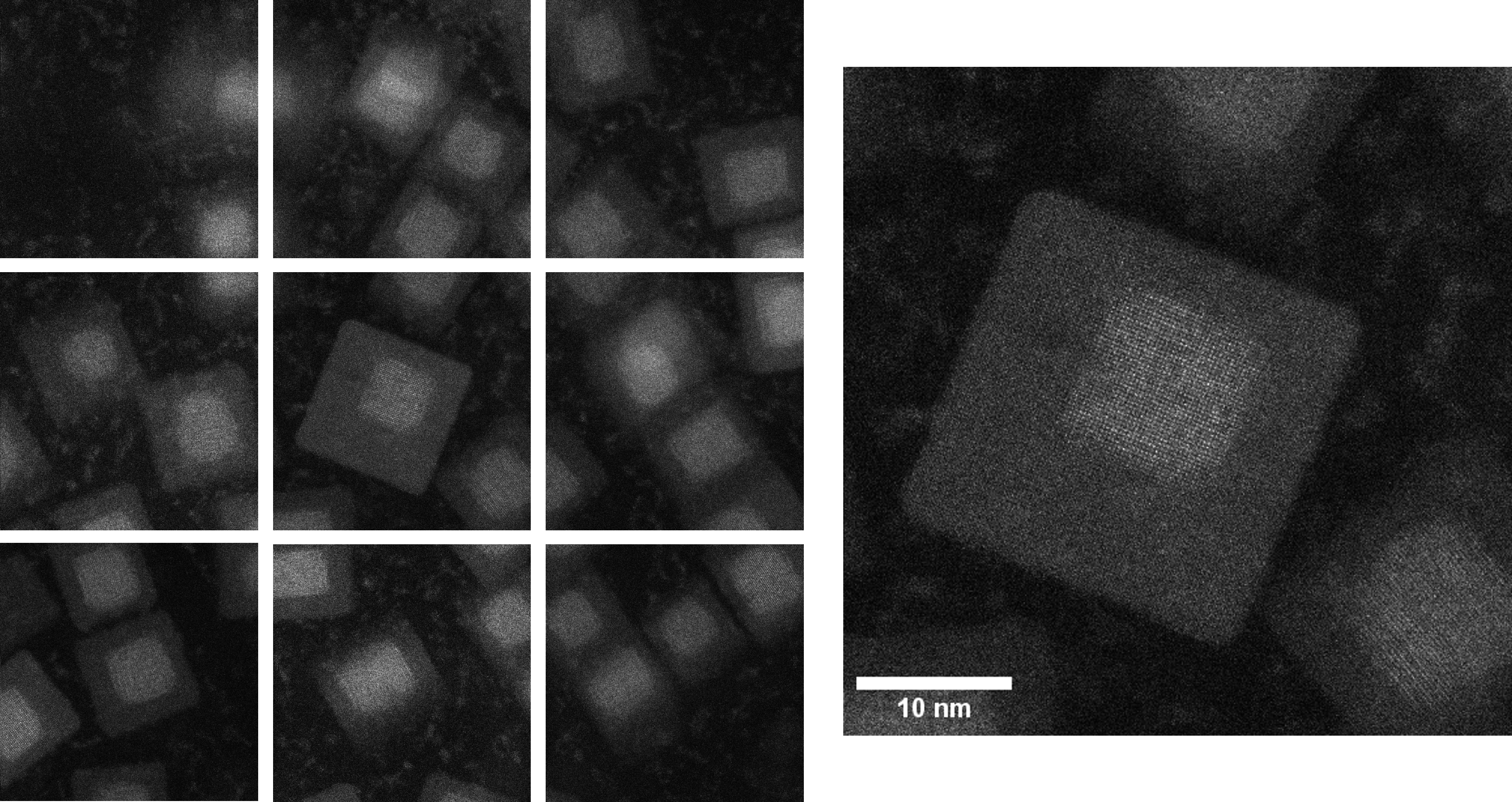}
\caption{Simultaneously acquired HAADF-STEM images for each stage position. The core/shell morphology and atomic structure is clearly visible due to the Z-contrast of this technique.}
\label{fig:adf grid}
\end{figure}

The TEAM 0.5 is also equipped with an advanced direct electron detector capable of acquiring diffraction patterns at 87,000 Hz for 4D-STEM \cite{Ercius2023-dz}. We integrated this camera with the automation system to enable multimodal data acquisition of phase contrast and dark field STEM images. Fig \ref{fig:HTP Client} shows a 3 by 3 grid of 4D-STEM datasets with 512 by 512 scan positions taken at 1.8Mx STEM magnification and a camera length of 170 mm (see Appendix \ref{sec:pipeline} for the full pipeline). The experiment took about 10 minutes and generated over 1.2 terabytes of raw data. The sample is a dense array of core/shell SrYbF$_5$/CaF$_2$ metal fluoride nanoparticles useful as upconverting optical probes in biological imaging \cite{Fischer2020-em, Pedroso2021-wv}. We show the ability to measure the atomic structure of several of these dose-sensitive nanoparticles using differential phase contrast STEM by 4D-STEM. Figure \ref{fig:4d-pos} shows the initial overview image used for position generation and registration where the colored dots indicate the positions and order of acquisition (black to yellow). Each data acquisition includes a HAADF-STEM image (Figure \ref{fig:adf grid}) and a 4D-STEM scan acquired simultaneously. The 4D-STEM data was analyzed using the center of mass technique to create a differential phase contrast measurement \cite{Shibata2012-rv}. The core/shell structure is obviously visible in the HAADF-STEM data due to the Z-contrast between the lighter shell Ca and heavier SrYb core; however, the weakly scattering F atoms are not resolved. DPC-STEM is sensitive to these weakly scattering atomic columns and are imaged in the center nanoparticle which is on the [100] zone axis. Automation allowed us to image several nanoparticles with advanced methods and minimal dose.

\begin{figure} [!htb]
\centering
\includegraphics [scale=0.45] {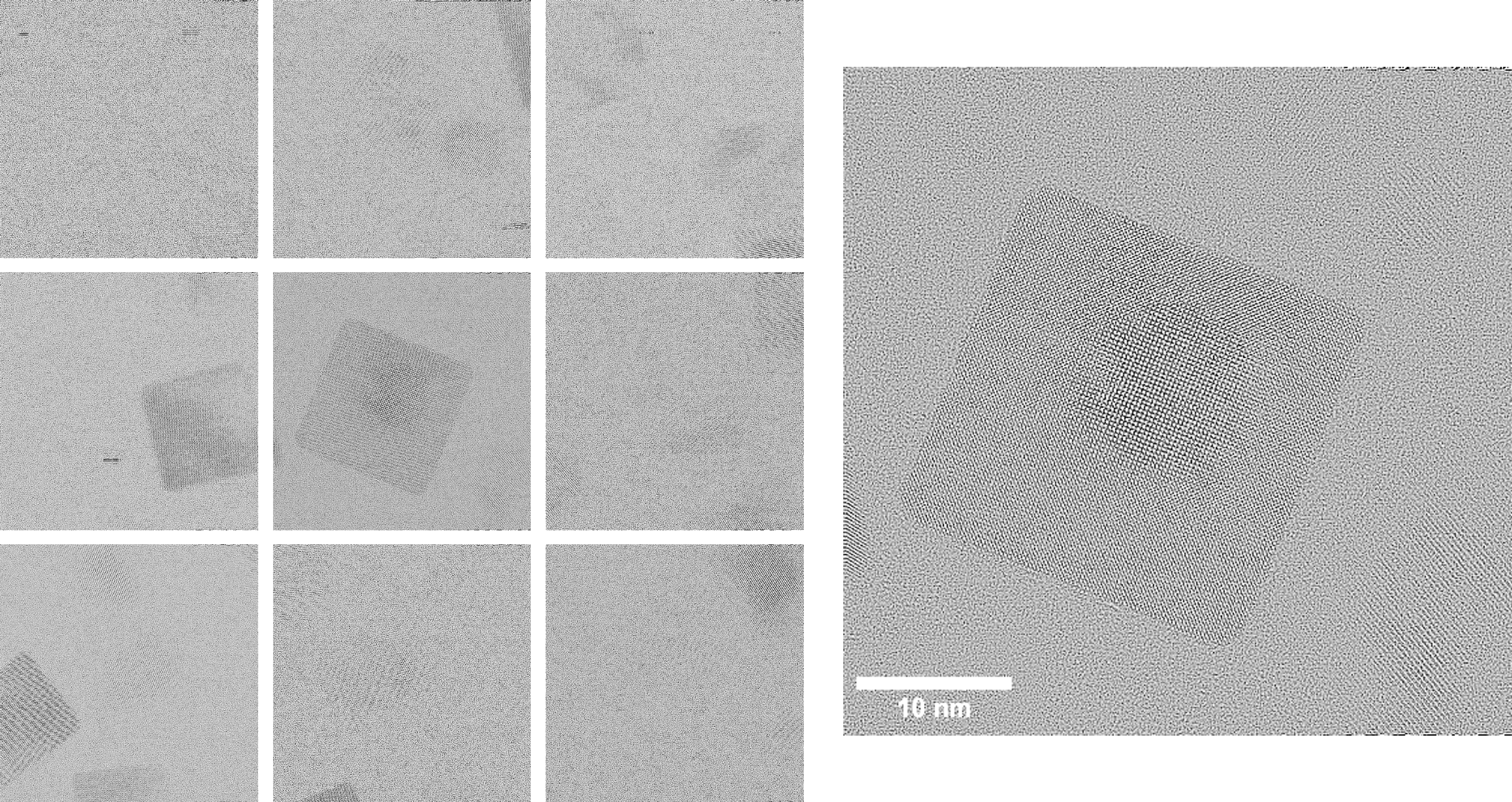}
\caption{The differential phase contrast STEM signal extracted from each 4D-STEM data set (140 Gigabytes each) acquired at each stage position. The lightly and heavily scattering atom columns are resolved in the on-axis nanoparticle in the center image. The center image is enlarged to show the available high-resolution structure allowing analysis of the core/shell interface.}
\label{fig:dpc grid}
\end{figure}

\section{Future Improvements}
The pipeline-based system described in this work provides a platform upon which future developments in STEM automation can be built. These could include more complex decision-making algorithms, potentially based on artificial intelligence and machine learning techniques, that would enable the system to make more intelligent choices regarding what data to take. One example of this is the need to image particles along certain zone axes. This could be accomplished by either identifying particles on-axis from a large population or automating the process of tilting them to the desired zone axis. The former could be implemented via image analysis techniques such as our FFT analysis similar of the CdSe/CdS experiment. The latter requires a far more complex system involving automating the determination of the tilts required from diffraction patterns and very accurate eucentric tilting.

\section{Summary}
We have implemented a pipeline-based system that utilizes the capabilities of the all-piezo TEAM stage for high-throughput acquisition of atomic-resolution HAADF and phase contrast STEM images on the TEAM 0.5 microscope. We have demonstrated the application of this system for the investigation of core/shell CdSe/CdS nanoparticles using HAADF-STEM and core/shell SrYbF/CaF nanoparticles using 4D-STEM. Alongside this, we have developed multiple techniques for STEM autofocusing and implemented them as part of this pipeline system. This system is available through the user program at the Molecular Foundry user center.

\section{Acknowledgements}

This work was partially funded by the US Department of Energy in the program ``4D Camera Distillery: From Massive Electron Microscopy Scattering Data to Useful Information with AI/ML." Work at the Molecular Foundry was supported by the Office of Science, Office of Basic Energy Sciences, of the U.S. Department of Energy under Contract No. DE-AC02-05CH11231. This research used resources of the National Energy Research Scientific Computing Center, a DOE Office of Science User Facility supported by the Office of Science of the U.S. Department of Energy under Contract No. DE-AC02-05CH11231. We would like to thank Gatan, Inc. as well as P Denes, A Minor, J Ciston, C Ophus, J Joseph, and I Johnson who contributed to the development of the 4D Camera.

\newpage
\appendix

\section{4D-STEM Pipeline}
\label{sec:pipeline}

The pipeline used for Fig \ref{fig:4d-pos}, \ref{fig:adf grid} and \ref{fig:dpc grid}
\singlespacing
\begin{verbatim}
1_start:
    tiling:
        nrperdim: (1, 1)
        mag: 320000
    focus:
        repeats: 1
    display:
        stitch: True
        panel: 0
    output:
        mag: 320000
        pixels: 1024
    stepper_attempts: 4
    return_to_start: True
    next_pipeline: 2_tiling
    
2_tiling:
    tiling:
        edge_tile_clearance: 1
        mag: 1800000
        stride: 1
    focus:
        repeats: 1
        nrslices: 9
        extra_focus_steps: [5.0e-09]
    registration:
        registration_end_mag: 1300000
        dfmax: 5e-07
        xytol: 5e-09
    display:
        panel: 1
    output:
        pixels: 512
        mag: 1800000
    errorskips: 10000
    chatmessageevery: 10
    next_pipeline: 3_record
    
3_record:
    record:
        dwell: 3e-06
        mag: 1800000
        pixels: 512
        zoomoutmag: 320000
        STEM camera length index: 10
        4D-STEM: True
    display:
        panel: 2
\end{verbatim}
\doublespacing

\section{CdSe/CdS Pipeline}
\label{sec:CdSSe Pipeline}
\singlespacing
\begin{verbatim}
1_start:
    tiling:
        nrperdim: (1, 1)
        mag: 5000
    focus:
        repeats: 3
    display:
        stitch: True
        panel: 0
    output:
        mag: 5000.0
        pixels: 1024
    stepper_attempts: 4
    return_to_start: True
    next_pipeline: 2_tiling

2_tiling:
    tiling:
        edge_tile_clearance: 2
        mag: 40000
        stride: 1
    registration:
        registration_start_mag: 10000
        registration_end_mag: 10000
        dfmax: 5e-07
        xytol: 2e-07
        focusatcenter: False
    focus:
        repeats: 3
    filter:
        passband: [5000, 12000]
        passfraction: 0.2
    display:
        panel: 1
    output:
        mag: 40000
        pixels: 1024
    stepper_attempts: 4
    next_pipeline: 3_tiling

3_tiling:
    tiling:
        edge_tile_clearance: 8
        mag: 640000
        stride: 1
    registration:
        registration_start_mag: 80000
        registration_end_mag: 160000
        dfmax: 5e-08
        xytol: 2e-08
        extra_focus_steps: [5e-09]
        focusatcenter: False
    filter:
        passband: [10000, 15000]
        passfraction: 0.5
    display:
        panel: 2
    output:
        pixels: 512
        mag: 640000
    errorskips: 10000
    chatmessageevery: 10
    next_pipeline: 4_record

4_record:
    record:
        dwell: 12e-06
        mag: 640000
        pixels: 2048
        zoomoutmag: 160000
    display:
        panel: 2
\end{verbatim}
\doublespacing

\newpage
\bibliographystyle{unsrt} 
\bibliography{Automation_References} 

\end{document}